\documentclass[runningheads]{llncs}
\usepackage{subfigure}
\usepackage{caption}
\usepackage{graphicx}
\usepackage{array}    
\usepackage{subfiles}
\usepackage{multirow}
\usepackage{helvet}
\usepackage{courier}
\usepackage{xcolor}
\usepackage{mathbbol}
\usepackage{enumitem}
\usepackage{graphicx}
\usepackage{booktabs}
\usepackage{amsmath}
\usepackage{algorithm}
\usepackage{algpseudocode}
\usepackage{relsize} 
\usepackage{wrapfig}
\usepackage{lipsum}  

\begin{document}
\newcommand{\holden}[1]{{\bf \color{magenta} [[$_{holden}$ ``#1'']]}}
\newcommand{\kmy}[1]{{\color{purple} [$_{Min}$ #1]}}
\newcommand{\reedit}[1]{{\color{black} #1}}

\title{Lightweight Modality Adaptation to Sequential Recommendation via Correlation Supervision}

\titlerunning{Lightweight Modality Adaptation to SR via Correlation Supervision}

\author{Hengchang Hu\inst{1} \and
Qijiong Liu\inst{2} \and
Chuang Li\inst{1} \and
Min-Yen Kan\inst{1}}

\authorrunning{Hu et al.}
\institute{National University of Singapore \\
\email{\{huhengc,lichuang,kanmy\}@comp.nus.edu.sg}
\and
The Hong Kong Polytechnic University\\
\email{jyonn.liu@connect.polyu.hk}\\
}


\maketitle             

\begin{abstract}

In Sequential Recommenders (SR), \textit{encoding} and \textit{utilizing} modalities in an end-to-end manner is costly in terms of modality encoder sizes.
Two-stage approaches can mitigate such concerns, but they suffer from poor performance due to modality forgetting, where the sequential objective overshadows modality representation.
We propose a lightweight knowledge distillation solution that preserves both merits: retaining modality information and maintaining high efficiency.
Specifically, we introduce a novel method that enhances the learning of embeddings in SR through the supervision of modality correlations. 
The supervision signals are distilled from the original modality representations, including both (1) holistic correlations, which quantify their overall associations, and (2) dissected correlation types, which refine their relationship facets (honing in on specific aspects like color or shape consistency).
To further address the issue of modality forgetting, we propose an asynchronous learning step, allowing the original information to be retained longer for training the representation learning module.
Our approach is compatible with various backbone architectures and outperforms the top baselines by 6.8\% on average.
We empirically demonstrate that preserving original feature associations from modality encoders significantly boosts task-specific recommendation adaptation. 
Additionally, we find that larger modality encoders (e.g., Large Language Models) contain richer feature sets which necessitate more fine-grained modeling to reach their full performance potential.

\end{abstract}

\section{Introduction}
\label{sec:intro}

Recommender systems reduce information overload by helping users select their preferred next items.
The Sequential Recommender (SR) paradigm specifically focuses on learning sequential relationships among the user's historical items, where the models range from GRU \cite{hidasi2015session}, Transformer \cite{kang2018self} to Graph Neural Network \cite{wu2019session} based methods.
The paradigm of using unique identities (IDs for short) to represent distinct items has been well-studied \cite{koren2009matrix,rendle2012bpr}.
Until recently, there has been a shift towards using modality to represent items \cite{yuan2023go}, with the emergence of sophisticated modality encoders such as Large Language Models (LLM) \cite{zeng2023glm130b,touvron2023llama} and large-scale image encoder models \cite{liu2021swin}.
These models represent modality features with vectors, discerning inherent modality correlations among items.

\begin{figure}[t]
    \centering
    \setlength{\abovecaptionskip}{-0cm}
    \setlength{\belowcaptionskip}{-0.6cm}\subfigure{\includegraphics[width=0.63\linewidth]{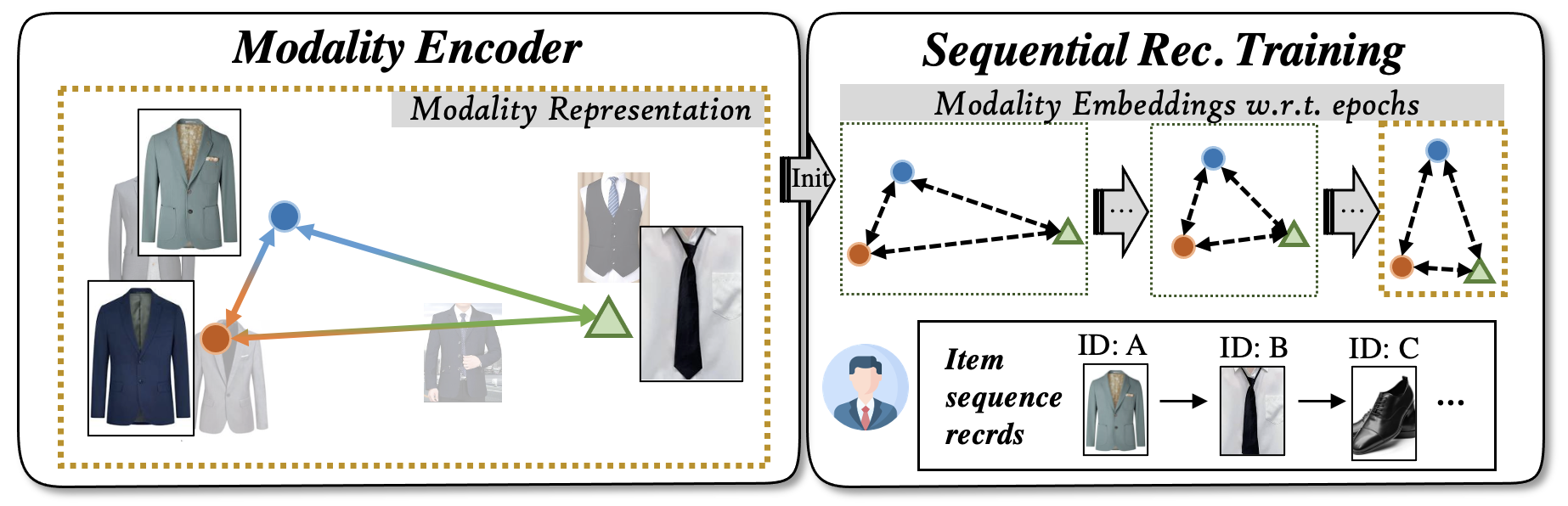}}
    \subfigure{
		\includegraphics[width=0.35\linewidth]{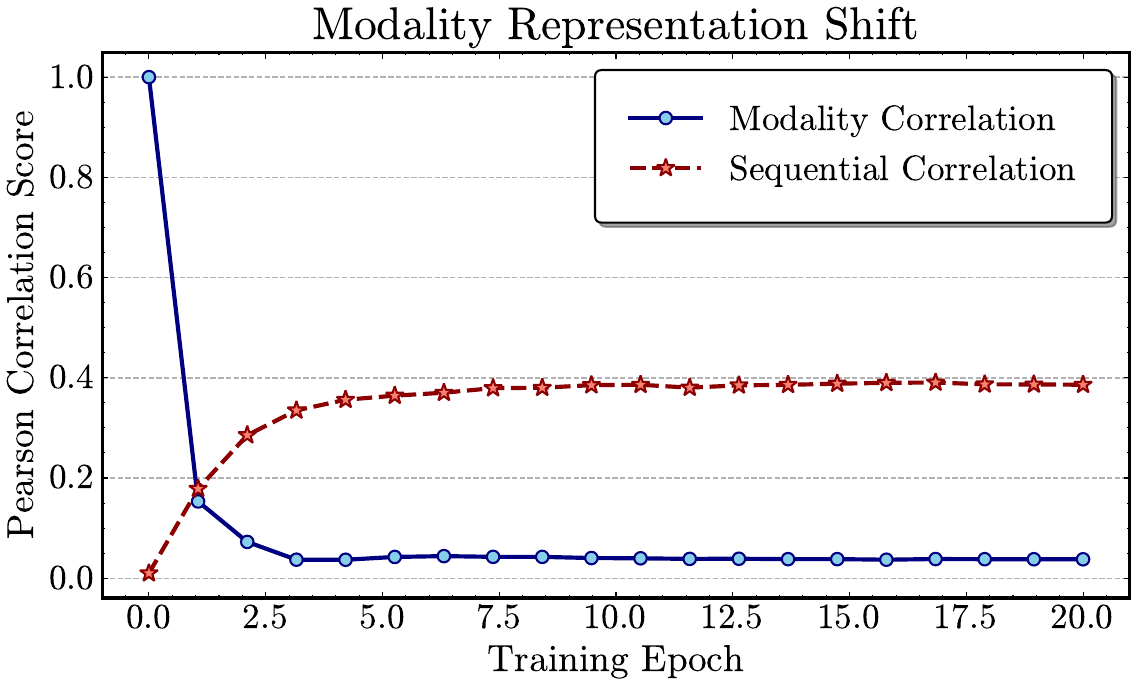}}
    \caption{(left) Diagram of adaptation from original modality representation to modality embedding over training epochs. (right) Experiment on the Beauty dataset with GRU-based models, highlighting the issue of modality forgetting.}
    \label{fig:intro}
\end{figure}

Nonetheless, directly utilizing encoded modality features to represent items presents a gap.
SRs are specifically designed to optimize item representations for \textit{sequential correlations} (i.e., selection order) modeling, making them ill-suited for leveraging the original modal representations (containing intrinsic \textit{modality correlations}) created by modality encoders.
Let us consider sequence selection in the fashion domain (Figure~\ref{fig:intro}): people 
usually pick a \texttt{suit} followed by a \texttt{tie}, rather than choosing two visually similar \texttt{suits}. 
Modality encoders can highlight their modality correlations (such as similarity), but they are easily overshadowed in SR training, as SR is tuned to capture item sequential relations. 

It is thus essential to tune and align modality knowledge to the SR systems.
While the end-to-end training of modality encoders and SR is a common practice \cite{elsayed2022end}, as it maintains the original features of items, it would be less feasible as the emergence of large-scale modality encoders such as LLMs.
Two-stage training \cite{mcauley2015image,covington2016deep} was proposed as a more efficient method that decouples the pretraining phase of the modality encoders and the downstream recommendation phase, which only takes cached modality representations as embedding input.

However, the embedding suffers from \textit{modality forgetting} in the recommendation phase.
This issue is demonstrated clearly in Figure~\ref{fig:intro} (right).
By training a modality-enriched SR model,
we compare the epoch-updated modality embeddings ($E$) with two other standard representation spaces: the original modality representations ($M$, including only modality correlations), and the item ID embeddings ($V$, that contains only sequential correlations) in a pure ID-based model.0
Assessing pairwise similarities in these spaces by Pearson score \cite{cohen2009pearson}, we find that
as training progresses, the item correlations within $E$ align more with $V$ but drift away from $M$ (the modality correlation rapidly deteriorates to nearly zero). 
As modality embeddings readily over-adapt to sequential correlations, it reduces their distinct advantage over plain IDs \cite{yuan2023go}.

Given the concerns above, we introduce a Knowledge Distillation \cite{hinton2015distilling} framework for modality-enriched SR (KDSR) that embraces both merits: it retains the modality original features from end-to-end training, while maintaining the effectiveness of two-stage training.
Specifically, it contains two components: a teacher model distills correlations from modality encoders, and a student model training embeddings supervised by these correlations.
We further propose designs to solve a few inherent challenges:
Since a unified correlation can only represent an overall relationship between two items, we design a differentiated holistic and dissected correlation operation to offer the modalities' interrelations capturing diverse perspectives, such as shape and color.
Furthermore, to allow for some discrepancies from noise, we employ a soft-matching mechanism to align teacher's and student's holistic correlations, 
and a vector quantization mechanism to convert dissected correlations into discrete codebook representations.
In this context, dissected correlations are transformed into a set of distinct \textit{codes}, simplifying complex relationships while retaining granular insights.

We conducted experiments across five distinct scenarios with image and text modalities. 
Our proposed KDSR effectively addresses the dilution of modal information, outperforming the top baselines by 6.8\%, on average.
Additionally, we introduce an asynchronous training step strategy that prevents the early loss of original modal details during training, further offering an enhanced adaptation. 
Our approach is compatible with different encoders and SR backbones.
Notably, our method integrates well with all large-scale modality encoders, like Swin \cite{liu2021swin} and LLaMA \cite{touvron2023llama}. 
\reedit{As these encoders expand in size, our quantization technique is able to yield deeper insights, with an adequate embedding dimension required to capture these more fine-grained correlations.}

Our contributions are summarized as:
(i) We are the first to spotlight rapid modal information dilution challenges, and propose a knowledge distillation framework to preserve the original modal information.
(ii) We offer a lightweight solution that combines coarse- and fine-grained correlation supervision in embedding learning. 
(iii) We conduct comprehensive experiments, which show our solution can 
keep pace with the evolving modality encoders, while being practical in real multi-modal use.

\section{Related Work}


\textbf{Modality-enriched recommender systems} typically utilize a modality encoder to represent item content knowledge, atop which a recommender model is applied for user modeling \cite{liu2019user} or collaborative filtering \cite{wei2019mmgcn}. 
One line of research is the joint training of the modality encoder and recommender~\cite{kang2017visually,oramas2017deep,wu2021empowering,liu2022multi}.
Another line takes a two-stage training paradigm \cite{hu2022modeling}, which firstly pretrains the modality encoder, and then makes extracted item representations detached from the modality encoder for downstream recommendation training. Such a two-stage design allows for more efficient use of resources, as large-scale modality encoders can be employed independently. Works in different domains (such as PREC~\cite{liu2022prec} in news recommendation, UVCAN~\cite{liu2019video} in micro-video recommendation, and CLEP~\cite{park2022exploiting} in music recommendation) have demonstrated the effectiveness of the application of extracted modality knowledge without updating the modality encoders. Our proposed KDSR method takes advantage of the efficient training of the two-stage design and the inherent modality correlation from the end-to-end paradigm.

\textbf{Knowledge distillation}~\cite{hinton2015distilling} is introduced to transfer valuable information from a large and complex model (i.e., ``teacher model'') to a smaller and more efficient one (``student''). Knowledge distillation techniques are categorized into supervised~\cite{liu2020general,zhao2022decoupled}, semi-supervised~\cite{zhou2020deep,su2021semi,cao2022semi} and self-supervised ones~\cite{lee2018self,rajasegaran2020self,xia2022device}.
It has been widely used in various domains, including computer vision~\cite{chen2022dearkd,gao2018image}, natural language processing~\cite{sanh2019distilbert,jiao2019tinybert}, and speech~\cite{lee2022fithubert,liu2019end}. In the recommendation domain, Tan {\it et al.} ~\cite{tan2016improved} first introduced knowledge distillation for SR. 
A portion of the previous work is aimed at improving efficiency~\cite{lian2020lightrec,chen2018adversarial}, while the majority is focused on enhancing inference quality~\cite{zhang2020distilling,xia2022device}.
However, an inadequately trained teacher can misguide, causing slow convergence leading to entrapment in local optima. In this paper, we design a self-supervised pipeline by introducing pre-calculated modality correlations.

\vspace{-0.1em}
\section{Methodology}
\vspace{-0.4em}

\textbf{Problem Formulation.}
Conventional sequential recommendation aims to predict the next item $v$ (from an overall set of $N$ items) that a user $u$ is likely to consume, based on their historical interactions $\{v_1, \dots, v_m\}$.
Extending this, Modality-enriched Sequential Recommendation (MSR) considers item modalities 
to refine the interrelationships among items within the sequence. 
Specifically, the objective is to acquire a function $f: \{(v_1,a_1,b_1), \dots, (v_m, a_m, b_m) \} \rightarrow v$,
where $a_i$/$b_i$ indicates the modality features (image/text) of item $v_i$ respectively. 

In this section, we present a novel method for optimizing modality usage in MSR via correlation supervision. 
We first outline the framework and then give the details of how we implement the correlation supervision in two phases: correlation distillation and approximation.

\vspace{-0.3em}
\subsection{Overall Framework of Knowledge Distillation for MSR}
\label{sec:pf_basemodel}
\vspace{-0.3em}

\begin{figure}[t]
    \setlength{\abovecaptionskip}{-0cm}
    \setlength{\belowcaptionskip}{-0.5cm}
    \centering
    \includegraphics[width=0.97\textwidth]{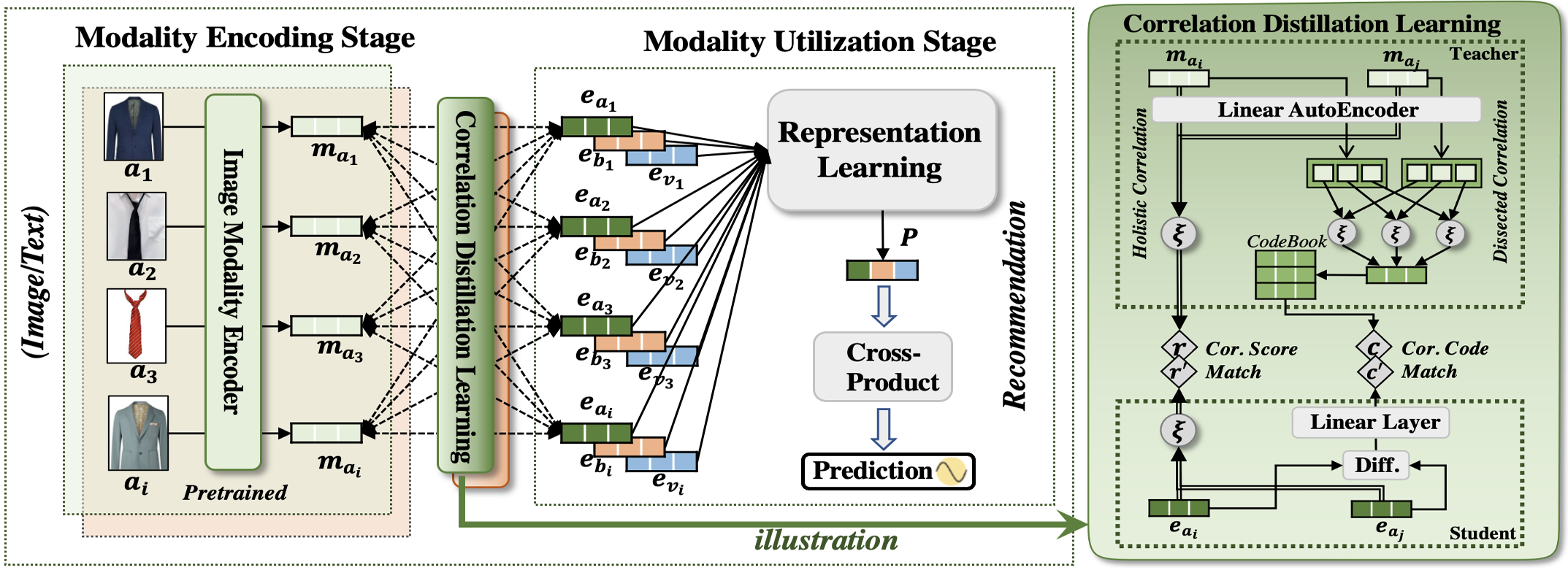}
    \caption{(left) Our KDSR framework. Green, orange, and blue denote image modality, text modality, and ID features, respectively. (right) Detail of Correlation Distillation Learning on the image (green) modality.}
    \label{fig:framework}
\end{figure}

Most existing MSR methods adhere to an Encoding--Utilization paradigm, 
as shown in Figure~\ref{fig:framework}. For simplification and clarity, we use the symbol $a$ in place of either $a$ or $b$ when referring to a specific modality type in the following text.

\textit{Feature Encoding.}
Item ID features $v$ are initially represented as integer values and can be converted into low-dimensional, dense real-value vectors $\mathbf{e}_v$ using a randomly initialized embedding table. 
Modality feature encoding comprises two key modules: (1) the feature representation module, which represents the original features with vectors $a \rightarrow \mathbf{m}_a$ via modality encoders \cite{he2016deep,raffel2020exploring} pretrained on extensive datasets; 
and (2) the embedding initialization module, which adapts these pretrained representations  $\mathbf{m}_a \rightarrow \mathbf{e}_a$ for recommendation tasks.

\textit{Feature Utilization.}
Several prior studies have focused on designing networks to model modality feature interactions and generate the overall sequence representation $\mathbf{P}$. Using the \textit{representation learning} function $g$, this is given by:
\begin{equation}
    \mathbf{P} = g(\mathbf{(e_{v_1},e_{a_1}, e_{b_1}), (e_{v_2},e_{a_2}, e_{b_2}),...,(e_{v_m},e_{a_m}, e_{b_m})})
\end{equation}

\textit{Prediction.}
To match the learned sequence representation $\mathbf{P}$ with candidate items $\mathbf{e}_{v}$, we adopt the wide-used dot product $\hat{y} = \mathbf{P} \cdot \mathbf{e}_{v}^\top$.
During training, the model measures the differences between the ground-truth $y$ and the predicted score $\hat{y}$ using the cross-entropy loss \cite{kang2018self}, denoted as $\mathcal{L}^{RS}(y,\hat{y})$.

\vspace{0.5em}
\noindent \textbf{Knowledge Distillation.}
As in the introduction, during training, $\mathbf{e}_a$ tends to deviate from $\mathbf{m}_a$, risking the loss of knowledge from the original feature $a$. 
To solve it, we employ a teacher--student paradigm for Correlation Distillation Learning (CDL), as illustrated in the central portion of Figure~\ref{fig:framework} (l).
The teacher model, during the feature encoding phase, distills correlations among item modalities and supervises the student model in the subsequent feature utilization stage. 
Formally, it predicts item pair correlations ($T(v_i, v_j) = r_{ij}$) relative to their modality representations, while the student model forecasts $S(v_i, v_j) = \hat{r}_{ij}$ using fine-tuned modality embeddings.
CDL's targets to minimize the disparity between $r_{ij}$ and $\hat{r}_{ij}$, which we detail next.


\subsection{Bridging the Gap: Correlation Distillation Learning}

As shown in Figure~\ref{fig:framework}~(right), we employ both coarse- and fine-grained supervision signals. The teacher distills both signals, and the student approximates them. \\

\noindent \textbf{Modality Correlation Distillation.}
The original $m_i$ derived from the modality encoder is often high-dimensional and laden with redundant data. For discerning refined and distinctive features from their vector representations, we employ a linear autoencoder $\mathbf{\tilde{m}}_i = AE(\mathbf{m}_{a_i})$ to convert $\mathbf{m}$ into a condensed $\mathbf{\tilde{m}}$ first to eliminate redundant features and focus on distinctive aspects.
Subsequently, CDL distills correlations into two representations. 

For example, using cosine similarity, the function calculates the angle's cosine between vectors $m_i$ and $m_j$, represented as $\xi(m_i,m_j)=\text{cos}(\theta_{m_i, m_j})$, emphasizing direction over magnitude.
Given the pivotal role of scoring functions in our study, we offer both distance and similarity-based methods for $\xi$. Their critical importance is highlighted in our experimental section.

\vspace{0.5em}
\noindent \textit{(1) Holistic} correlation scores $T_{s}: \xi(\mathbf{\tilde{m}}_i,\mathbf{\tilde{m}}_j) \rightarrow r_{ij}$.
We use an unsupervised method to determine the disparity between paired representations via the correlation scoring function \( \xi \), yielding a continuous value \( r \). For instance, with cosine similarity, \( \xi(\mathbf{m}_i, \mathbf{m}_j)=\text{cos}(\theta_{m_i, m_j}) \) computes the cosine of the angle between vectors \( \mathbf{m}_i \) and \( \mathbf{m}_j \), prioritizing direction over magnitude. 
Given the pivotal role of the scoring function in our study, 
our experimental results showcase the  significance of both distance and similarity-based approaches to $\xi$.

\vspace{0.5em}
\noindent \textit{(2) Dissected} correlation codes $T_c: \pi(\mathbf{\tilde{m}}_i,\mathbf{\tilde{m}}_j) \rightarrow c_{ij}$.
Considering just vector-wise operations $\xi$ may be overly simplistic;
we distill fine-grained discrete correlation signals \( c \) through a more detailed process \( \pi \). Given that modality vectors capture diverse features like color and shape, an in-depth approach is also vital to model their intrinsic relationships. For example, the pairs (\texttt{black-suit}, \texttt{black-tie}) and (\texttt{black-suit}, \texttt{yellow-suit}) might both exhibit high correlation, but their attributes differ. Therefore, pinpointing these unique correlation patterns into distinct codes is essential for such generalization. This process involves two main steps: correlation vectorization and vector quantization.

In the first step, rather than representing the correlation with a single scalar value $r_{ij}$, we vectorize their correlations in a vector $\mathbb{r}_{ij}$.
Drawing inspiration from Product Quantization \cite{jegou2010product}, we treat each sub-vector from $\mathbf{\tilde{m}}$ as an individual, meaningful segment (where $\mathbf{\tilde{m}}$ from the autoencoder offers distinct, significant features).
Specifically, we split $\mathbf{\tilde{m}}_i$ into $D$ sub-vectors, represented as $[\mathbf{\tilde{m}}_{i,1}; ...; \mathbf{\tilde{m}}_{i,D}]$. Then we use the scoring function $\xi$ to measure the correlations for each sub-vector in $\mathbf{\tilde{m}}_{i}$ against its counterpart in $\mathbf{\tilde{m}}_{j}$. These scores are then concatenated to form correlation vectors as $\mathbb{r}_{ij} = [\xi(\mathbf{\tilde{m}}_{i,1},\mathbf{\tilde{m}}_{j,1}); ...; \xi(\mathbf{\tilde{m}}_{i,D},\mathbf{\tilde{m}}_{j,D})]$. 
As such, it provides a more nuanced set of relationships between pairs of sub-vectors within the vectors.

In the second step, to condense the information and encapsulate the correlation pattern, we use Vector Quantization (VQ) \cite{hou2023learning} to convert the correlation vectors $\mathbb{r}_{ij}$ into correlation codes $c_{ij}$. 
This approach retains signal while accommodating for minor errors in the imperfect alignment between the encoder's modality spaces and the SR embedding space.
We start by generating a codebook composed of $x$ vectors, where $x$ is a predetermined number of codes. Each \textit{codeword} in this codebook denotes a distinct correlation type. They are initialized randomly, and then optimized with training data to better match the vector distribution of $\mathbb{r}$.
Each vector is then mapped to its nearest codeword, based on Euclidean distance (a common practice for quantized code learning \cite{hou2023learning}).
This process transforms $\mathbb{r} \in \mathbb{R}^{N \times N \times D}$ to $c \in \mathbb{R}^{N \times N \times 1}$, where each entry signifies an index value between 1 and $x$, representing the closest codeword. 
While there are other quantization methods which we tried (see \S\ref{ss:mcs}), this method validated well empirically.

With the aforementioned steps, similar correlation patterns are mapped to the same codeword. This process also distinguishes subtle patterns, such as \textit{``similar color, different shape"} versus \textit{``similar shape, different color"}, through distinct codewords, highlighting more nuanced relationships. \\

\noindent \textbf{Distilled Correlation Approximation.}
To ensure correlation consistency between embeddings $E$ in the modality utilization stage and modality representation $M$ in the modality encoding stage, the student network needs to mirror the outcomes of the more complex teacher model. Specifically, we assess item correlation through holistic proximity and directional relations, which directly pair with holistic correlation scores $T_s$ and dissected correlation codes $T_c$. 

\vspace{0.5em}
\noindent (1) \textit{Holistic} correlation learning.
In the student model, the correlation between items \( v_i \) and \( v_j \) is derived from their fine-tuned embeddings $\mathbf{e}_{a_{i}}$ and $\mathbf{e}_{a_{j}}$:
\begin{equation}
    \hat{r}_{ij} = \xi \left(g_\phi (\mathbf{e}_{a_{i}}), g_\phi (\mathbf{e}_{a_{j}}) \right)  
\end{equation}
where $g_\phi$ is a transformation linear layer.
In recommendation settings, item similarity is not as crucial as in image classification tasks. Thus, we adjust the sensitivity of modality supervision by tweaking the temperature parameter $\tau$ in a soft match equation, where the loss function is defined as:
\begin{equation}
    \mathcal{L}^{K D}_{S}=\frac{1}{N^2} \textstyle\sum_{i=1,j=1}^{N,N} ||\sigma\left(r_{ij} / \tau\right), \sigma\left(\hat{r}_{ij} / \tau\right)||_{l2}
\end{equation}
Here $r$ and $r'$ are predictions from the teacher and student models, respectively, and $l2$ is the squared error loss. Raising $\tau$ smooths the probability distribution, reducing disparities. Conversely, decreasing the temperature sharpens the distribution, amplifying the distinctions between probabilities. 

\vspace{0.5em}
\noindent (2) \textit{Dissected} correlation learning.
In the teacher model, VQ is employed to differentiate correlation patterns.  However, it also incurs increased complexity. Alternatively, we use a simpler, computationally-efficient direct subtraction mechanism to simulate inter-modal correlation patterns in student models; i.e., terming different correlation facets as directional relations in embedding spaces. Specifically, for any item pair, the student model first determines their directional relationship via a subtraction vector, subsequently predicting a probability distribution $p$ across $x$ classes. This prediction logit $p$ aligns with the correlation code probabilities supplied by the teacher model.
Mathematically,
\begin{equation}
    \tilde{e}_{ij} = |\mathbf{e}_{a_{i}} - \mathbf{e}_{a_{j}}|, \text{~~} p = W_{out}(ReLU(W_{c}(\tilde{e}_{ij})))
\end{equation}
Here, $W_{c} \in \mathcal{R}^{d \times d}$ and $W_{out} \in \mathcal{R}^{d \times x}$ are two linear transformation weight matrices, and \textit{ReLU} is the activation function.
The teacher's output $c_{ij}$ is expressed as a binary distribution, where the index of the predicted correlation code corresponds to 1; otherwise, 0.
The loss function for learning can then be formulated as the cross-entropy loss between the student's and teacher's output distributions:
\begin{equation}
    \mathcal{L}^{K D}_{C}= -\frac{1}{N^2} \textstyle\sum_{i=1,j=1}^{N,N} \textstyle\sum_{t=1}^x 1(c_{ij}=t) \log \left(p_{ij, t}\right)
\end{equation}


\subsection{Prediction \& Training Strategy}

As depicted in Figure~\ref{fig:framework}, sequences of image, text, and ID embeddings---denoted as \(\{\mathbf{e}_{a_i}\}\), \(\{\mathbf{e}_{b_i}\}\), and \(\{\mathbf{e}_{v_i}\}\)---are channeled into representation learning modules. 
For instance, we utilize MMSR \cite{hu2023adaptive} as our representation learning module, leveraging a heterogeneous GNN for information exchange.
Initializing \(\mathbf{e}_a\), \(\mathbf{e}_b\), and \(\mathbf{e}_v\) as distinct node types and using sequential transitions as edges, we obtain a sequence of hidden states \([\mathbf{h}_1, ..., \mathbf{h}_m]\) after $l$ layers of graph propagation.

The final sequence representation \(\mathbf{P}\) integrates both long- and short-term interests. For the long-term ones, the mean is calculated as 
\(h^l = avg(\mathbf{h}_i)_{i=1}^{m}\). For the short-term interests, we concentrate on the sequence's last \(q\) items in the sequence, applying attention mechanisms. The final \(\mathbf{P}\) is calculated by:
\begin{equation}
    \alpha_i = W_2(ReLU(W_1 \mathbf{h}_{i})), \text{~}
    h^s = \textstyle\sum_{i=m-q+1}^{m} (\alpha_i \mathbf{h}_i) 
    \text{, and}\quad
    \mathbf{P} = W_3[h^s; h^l],
\label{eq:short-seq-h}
\end{equation}
with transformation matrices \(W_1 \in \mathbb{R}^{d \times d}\) and \(W_2 \in \mathbb{R}^{d \times 1}\), the attention weights, \(\alpha \in \mathbb{R}^{q \times 1}\), emphasize recent items. Combining short-term interest \(h^s\) with long-term \(h^l\) ones yields the final sequence representation 
\(\mathbf{P}\) and the matrix \(W_3 \in \mathbb{R}^{2d \times d}\) projects this combined representation.

Finally, to supervise the embedding learning using correlation, we combine the supervision loss with prediction loss, resulting in the training objective:
\begin{equation}
    \mathcal{L} = \mathcal{L}^{RS}(y,\hat{y}) + \lambda_1 \mathcal{L}^{KD}_S + \lambda_2 \mathcal{L}^{KD}_C
\end{equation}

\noindent \textbf{Asynchronous Training.}
We also explore adaptation operations that prioritize preserving the original domain's information while training new domain parameters. 
Specifically, we use an asynchronous update training strategy \cite{zhang2015deep} for modality embedding parameters and other representation learning parameters.
The learning rates for embedding parameters begin at \(0.1\epsilon\) and linearly increase to \(\epsilon\) over the first $\eta$ epochs, with \(\epsilon\) being the learning rate for other parameters and $\eta$ is the pre-set parameter for stopping the asynchronous training.
The effectiveness of this training strategy is validated in our ablation study, detailed in \S~\ref{ss:mcs}.

\vspace{-0.25em}
\subsection{Complexity Analysis}
\vspace{-0.15em}

Compared to standard two-stage methods, our approach comes with the additional cost of  CDL. 
Using efficient scoring functions like cosine similarity (costing \(O(d)\) with \(d\)-dimensional vectors for each item pair), CDL adds a cost of \(O(L^2d)\) per sequence during training, where $L$ denotes the average sequence length. Compared to backbones such as SASRec, which uses transformer cores and requires \(O(L^2dH)\) 
(with $H$ being the number of the attention head) 
for representation learning, CDL adds an additional computational increase of $1/(1+H)$, largely negligible compared to end-to-end training costs.
In end-to-end training, the main cost comes from tuning the modality encoder, which has a very large parameter size (seen in \S~\ref{sec:exp-3}), compared to the parameter size for training SR.
During the inference phase, the student model operates independently, so its computational complexity is \textbf{the same} as standard two-stage models.

\vspace{-0.2em}
\section{Experiment}
\vspace{-0.45em}

\noindent \textbf{Datasets.}
Consistent with prior MSR work \cite{he2016vbpr,zhang2021mining}, we used the Amazon review \cite{he2016ups} and MovieLens \cite{harper2015movielens} datasets for our experiments, both of which provide item descriptions and images.
For Amazon data, to ensure our approach's versatility in various scenarios, we  selected four broad per-category datasets, namely Beauty, Clothing, Sport, and Toys. 
In these datasets, each review rating (of products or movies) signifies a positive user--item interaction, as commonly accepted in prior studies \cite{he2016vbpr,he2020lightgcn,zhang2021mining}. To ensure a fair comparison with existing methods, we implemented core-5 filtering, a method that iteratively filters the dataset until each user and item has at least five interaction records.

\begin{table}[h]
\scriptsize
\centering
\renewcommand{\arraystretch}{1.1}
\setlength\tabcolsep{4.2pt}
\setlength{\belowcaptionskip}{-2em}
\begin{tabular}{l|c|c|c|c|c}
\hline 
 & \textbf{Beauty} & \textbf{Clothing} & \textbf{Sports} & \textbf{Toys} & \textbf{ML-1M} \\ \hline
\textbf{User \#}  & 22,363  & 39,387 & 35,598 & 19,412 & 6,040 \\ 
\textbf{Item \#}  & {12,101} & {23,033} & {18,357} & {11,924}  & 3,416   \\ 
\textbf{Inter. \#}  & {198,502} & {278,677} & {296,337} & {167,597} & 999,611  \\ 
\textbf{Avg Len. \#}  & {8.88}   & {7.12}     & {8.46}   & {8.79} & 165.50  \\ 
\textbf{Sparsity}  & {99.93\%}   & {99.97\%}     & {99.95\%}   & {99.93\%} & 95.16\% \\ \hline
\end{tabular}
\caption{Dataset statistics after preprocessing. }
\label{tab:datasets}
\end{table}


\noindent \textbf{Baseline Models.}
We select three baseline categories for comparison.
\textbf{(A)} ID-based SR models: GRU4Rec \cite{hidasi2015session}, using Gated Recurrent Units; SASRec \cite{kang2018self}, employing self-attention; and SR-GNN \cite{wu2019session}, a graph-based model considering user-item and item-item relations.
\textbf{(B)} MSR models with frozen features: MMSRec \cite{song2023self}, which uses features as supervised signals, and modified SASRec models (SASRM) integrating static modality and ID features for representation learning.
\textbf{(C)} MSR models with fine-tuned features: NOVA \cite{liu2021noninvasive} and DIF-SR \cite{xie2022decoupled}, applying non-invasive fusion methods, and Trans2D \cite{singer2022sequential} and MMSR \cite{hu2023adaptive}, which employ holistic fusion to merge items' modality and ID features.

\vspace{0.3em}
\noindent \textbf{Evaluation Protocol.}
Following conventional SR evaluations, we split each user's sequence: the first 80\% for training and the last 20\% for testing, where each user has at least one test data point. 
Performance is gauged using two common ranking metrics: hit ratio (HR, or H@$k$) and mean reciprocal rank (MRR, or M@$k$), with higher values denoting superior efficacy in top $k$ predictions.

\vspace{0.3em}
\noindent \textbf{Parameter Settings.}
For a fair comparison, all models use a 128-dimensional embedding and 512 batch size. Learning rates spanned logarithmically from $10^{-1}$ to $10^{-4}$, and $L_2$ regularization ranged from $\{0, 10^{-5}\}$. Dropout ratios were set from 0 to 0.9. The Adam optimizer \cite{kingma2014adam} was used.
Each test was averaged over five runs for consistency. $\tau$ is set to 1.5. We set the optimized quantization split $D=8$ and code number $x=100$. We adjusted the ($\lambda_1$,$\lambda_2$) settings to (1, 0.5) for the Amazon dataset and (0.5, 0.1) for the ML-1M dataset. 

\subsection{Performance Comparison}

As seen in Table~\ref{tab:result}, even when leveraging the best baseline as the representation learning module, our KDSR consistently surpasses top benchmarks.

Among ID-based models, SASRec stands out, emphasizing the effectiveness of attention mechanisms. 
Among MSR models incorporating modality features, the fine-tuned modality approach (MMSR) works better on datasets with shorter sequences (like Beauty, Sport, and Toy). However, for longer sequences as in the ML-1M dataset, the frozen modality method (SASRM) prevails. 
This underscores that the fine-tuned approach can capture the inherent modal information in shorter sequences, but maintaining this information becomes problematic in longer sequences, rendering the frozen feature approach more favorable.
Meanwhile, in fusion-based models, holistic methods \cite{hu2023adaptive} like MMSR or Trans2D outperform early/late fusion approaches \cite{hu2023adaptive} like NOVE and DIF-SR on shorter Amazon datasets, with MMSR being particularly effective. However, for longer sequences, the advantage shifts away from these holistic methods. 
This indicates that fusing modality information into items \textit{during} sequential modeling benefits shorter sequences, but it can complicate and diminish results for longer ones.

For our KDSR, there was a notable increase in HR (average of $\uparrow$10.2\%), compared to a modest rise in MRR (average of $\uparrow$3.4\%). This was particularly evident in the Sport and ML-1M datasets. The difference indicates that while harnessing original modality correlation can boost recommendation overall precision (i.e., HR), its influence on ranking improvement (i.e., MRR) might be limited.

\begin{table*}[t]
\renewcommand{\arraystretch}{0.95}
\setlength\tabcolsep{0.9pt}
\setlength{\belowcaptionskip}{-1em}
\scriptsize
\begin{tabular}{c|l|ccc|cc|cccc|c}
\toprule
    & Metric  & GRU4Rec & SASRec & SR-GNN & MMSRec & SASRM & NOVA & DIF-SR & Trans2D & MMSR & KDSR   \\ \hline \midrule
\multirow{4}{*}{\rotatebox{90}{\textbf{Beauty}}}   
     & H@5 & 5.6420 & 6.1900 & 4.1483           & 5.2851 & 6.3078            & 4.2219 & 6.5789 & 6.0191          & \underline{7.1563} & \textbf{7.4156}$^{\dagger\star}$  \\
     & M@5 & 3.1110 & 3.2165 & 2.2123           & 3.0391 & 3.5983            & 2.1785 & 4.0735 & 3.4387          & \underline{4.4429} & \textbf{4.6355}$^{\dagger\star}$ \\
     & H@20 & 12.7217 & 14.0681 & 10.2351       & 11.9568 & 13.7529            & 10.7978 & 14.0137 & 13.2214         & \underline{14.1470} & \textbf{14.675}$^{\dagger\star}$ \\
     & M@20 & 3.7714 & 3.9668 & 2.7911           & 3.6539 & 4.2998            & 2.8160 & 4.7983 & 3.9460          & \underline{5.0433} & \textbf{5.2444}$^{\dagger\star}$ \\
\midrule
\multirow{4}{*}{\rotatebox{90}{\textbf{Cloth.}}}     
     & H@5 & 1.3340 & 1.5885 & 0.8547           & 1.3995 & 1.9660            & 1.2937 & 1.5524 & 1.3929          & \underline{1.8684} & \textbf{2.4368}$^{\dagger\star}$ \\
     & M@5 & 0.6765 & 0.7820 & 0.4555           & 0.7841 & 0.9992            & 0.6503 & 0.7961 & 0.6682          & \underline{1.1365} & \textbf{1.1647}$^{\dagger\star}$  \\
     & H@20 & 3.8111 & 3.9574 & 2.7528           & 3.8329 & 4.8163            & 3.4866 & 4.0571 & 4.0683          & \underline{4.4136} & \textbf{4.9499}$^{\dagger\star}$  \\
     & M@20 & 0.9418 & 1.0339 & 0.6251           & 1.0384 & 1.2683            & 0.8783 & 1.0530 & 1.0391          & \underline{1.3344} & \textbf{1.3851}$^{\dagger\star}$ \\
\midrule
\multirow{4}{*}{\rotatebox{90}{\textbf{Sport}}}     
     & H@5 & 2.4388 & 2.9549 & 2.0742           & 2.8996 & 3.0306            & 2.1539 & 2.5145 & 2.7168          & \underline{3.2657} & \textbf{3.5008}$^{\dagger\star}$ \\
     & M@5 & 1.2696 & 1.5858 & 1.0790           & 1.5391 & 1.6625            & 1.1271 & 1.3469 & 1.4235          & \underline{1.9846} & \textbf{1.9934}$^{\star}$  \\
     & H@20 & 6.6430 & 7.2208 & 5.4376           & 6.8258 & 7.3683            & 5.8062 & 7.0774 & 6.9453          & \underline{7.7466} & \textbf{7.8126}$^{\dagger\star}$ \\
     & M@20 & 1.6947 & 2.0357 & 1.4349           & 1.8650 & 2.0682            & 1.5648 & 1.9214 & 1.7058          & \underline{2.2826} & \textbf{2.3066}$^{\star}$ \\
\midrule
\multirow{4}{*}{\rotatebox{90}{\textbf{Toys}}}    
    & H@5 & 3.8663 & 5.0902 & 2.7329           & 4.1151 & 4.6208            & 3.7899 & 5.2363 & 4.1908          & \underline{6.1159} & \textbf{6.6027}$^{\dagger\star}$ \\
    & M@5 & 2.0022 & 2.7536 & 1.4878           & 2.5188 & 2.7823            & 1.9641 & 3.1944 & 2.2370          & \underline{3.8987} & \textbf{4.2785}$^{\dagger\star}$ \\
    & H@20 & 10.0727 & 11.8668 & 6.7452        & 9.9061 & 9.5824            & 9.0609 & 12.0284 & 10.5082          & \underline{12.1192} & \textbf{12.3361}$^{\dagger\star}$ \\
    & M@20 & 2.7267 & 3.4228 & 1.8655           & 3.0021 & 3.2572            & 2.4502 & 3.8777 & 2.9298          & \underline{4.3551} & \textbf{4.7925}$^{\dagger\star}$ \\
\midrule
\multirow{4}{*}{\rotatebox{90}{\textbf{ML-1M}}}  
    & H@5 & 12.0854 & 10.6164 & 10.0049        & 14.0942 & \underline{15.4602} & 13.8782 & 14.2002 & 13.2905          & 12.5701 & \textbf{15.5644}$^{\dagger\diamond}$ \\
    & M@5 & 6.3590 & 5.1311 & 5.1460          & 7.6364 & \underline{8.8079} & 7.5799 & 7.4881 & 7.1383          & 6.6671 & \textbf{8.8109}$^{\dagger\diamond}$ \\
    & H@20 & 29.3807 & 29.1878 & 25.8203       & 31.2147 & \underline{32.9212} & 31.2121 & 33.1632 & 30.844         & 30.0923 & \textbf{33.1966}$^{\dagger\diamond}$ \\
    & M@20 & 7.9991 & 6.8462 & 6.6340         & 9.4173 & \underline{10.4825} & 9.2362 & 9.3023 & 8.3520         & 8.3287 & \textbf{10.4995}$^{\dagger\diamond}$ \\
\bottomrule
\end{tabular}
\caption{
Overall performance (\%). Bold denotes peak average; underline highlights best baselines. $\dagger$ marks KDSR's statistical significance ($p$-value$<0.05$) against the best baseline. KDSR builds on the best per-dataset baseline as the backbone, marked as $\star$ (MMSR) or $\diamond$ (SASRM).
}
\label{tab:result}
\end{table*}

\begin{figure}[t!]
    \centering
    \setlength{\abovecaptionskip}{-0.1em}
    \setlength{\belowcaptionskip}{-2em}
    \subfigure{
		\includegraphics[width=0.315\linewidth]{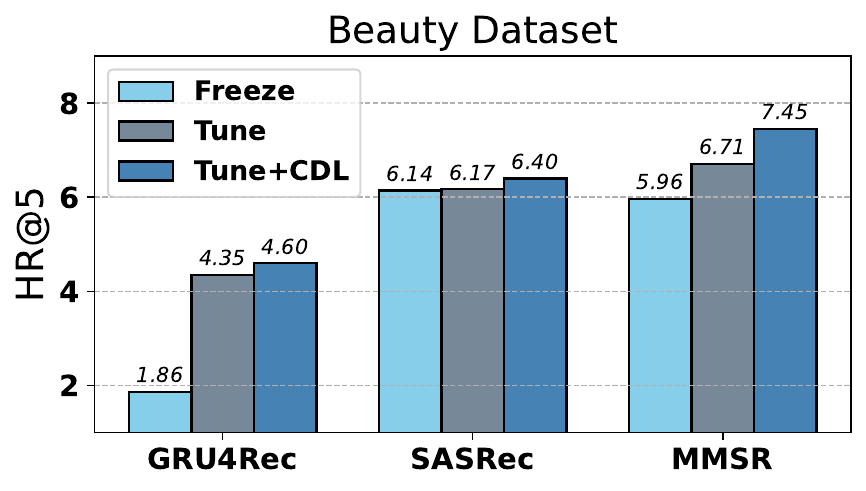}}
    \subfigure{
		\includegraphics[width=0.315\linewidth]{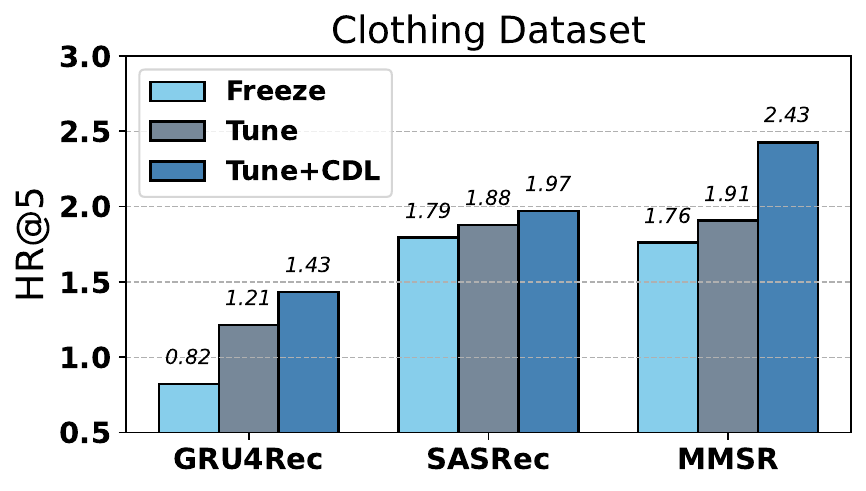}}
    \subfigure{
		\includegraphics[width=0.315\linewidth]{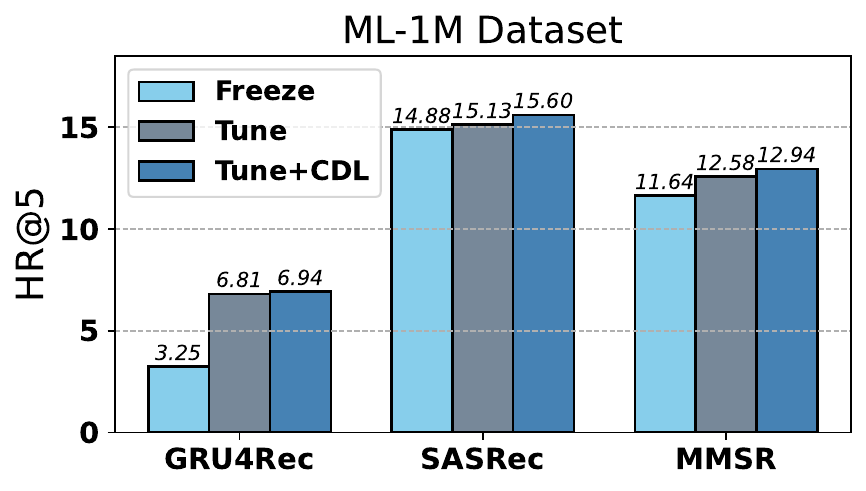}}
    \caption{Compatibility study with different representation learning backbones.}
    \label{fig:backbone}
\end{figure}

\vspace{0.5em}
\noindent \textbf{Compatibility Study.}
Observing that optimal representation varied across learning backbones for tasks with long and short sequences, we further explored CDL's compatibility with different backbones, as shown in Figure~\ref{fig:backbone}.


Across the three \textit{representation learning} backbones, a clear trend emerges: 
\textit{frozen} embeddings are less effective than \textit{tuning}, and our CDL enhances embedding tuning adaptability. For GRU4Rec, frozen embeddings yield notably lower scores, given the recurrent models' demand for more adaptive embedding inputs.
In contrast, SASRec, which concentrates on using attention scores to signify pair-wise correlations among items, exhibits the least fluctuation in training strategies among the three backbones, as even original representations yield meaningful attention calculation. 
MMSR shines on the Clothing dataset with the shortest average sequence, but falters with longer sequences from ML-1M.  
In ML-1M, although all backbones improve slightly with CDL, the gains are modest. This could be because supervising correlations in lengthy sequences may apply overly tight constraints, considering the exponential growth of item correlation pairs with sequence length, which in turn can impede effective embedding learning in recommendation systems.


\subsection{Modality Correlation Study}
\label{ss:mcs}
\noindent \textbf{Ablation within the CDL module.}
To investigate the efficacy of components in the student model, we conducted an ablation study, as illustrated in Table~\ref{tab:ablation}.

The soft match in KDSR, when compared to a hard match (w/o $\tau$), shows improved performance. This improvement is more evident in H@20 but less in H@5, especially in the Beauty dataset where the soft match actually decreases performance. This suggests that a hard match might compromise recommendation diversity, increasing overall hit rates but negatively affecting the top recommendations.
Using single correlation supervision signals, the holistic correlation score signal (w/o $\mathcal{L}^{KD}_C$) outperforms the dissected correlation code (w/o $\mathcal{L}^{KD}_S$). 
Yet, combining dissected and holistic signals further enhances performance compared to using holistic ones solely, particularly in H@20. However, there is a slight drop in H@5 for the Sports dataset. This implies that holistic signals mainly boost recommendation accuracy, while dissected signals add to diversity, as more fine-grained guidance. 
As diversity increases, performance under broader top rankings will increase more.
Last but not least, omitting asynchronous training (AT) or the dual supervision signals mildly impacts performance.  We believe they have interdependent roles in preserving the original modality information.

\begin{table}[t!]
\setlength{\belowcaptionskip}{-2em}
  \begin{minipage}[t]{0.57\textwidth}
    \vspace{0pt}
    \centering
    \scriptsize
    \tabcolsep=0.9mm
    \begin{tabular}{l|cc|cc|cc}
    \toprule
    \multirow{2}{*}{Model} & \multicolumn{2}{|c}{\textbf{Beauty}} & \multicolumn{2}{|c}{\textbf{Clothing}} & \multicolumn{2}{|c}{\textbf{Sport}}  \\
     & H@5 & H@20 & H@5 & H@20 & H@5 & H@20    \\
    \midrule
    \textbf{KDSR} & 7.416 & 14.675 & 2.437 & 4.950 & 3.501 & 7.813 \\
    \textit{w/o} $\tau$ & 7.460 & 14.340 & 2.369 & 4.861 & 3.466 & 7.422 \\
    \textit{w/o} $\mathcal{L}^{KD}_C$ & 7.407 & 14.292 & 2.367 & 4.860 & 3.470 & 7.641 \\
    \textit{w/o} $\mathcal{L}^{KD}_S$ & 7.395 & 14.215 & 2.291 & 4.853 & 3.422 & 7.715 \\
    \textit{w/o} $\mathcal{L}^{KD}_{C\&S}$ & 7.045 & 13.951 & 2.269 & 4.716 & 3.391 & 7.623 \\
    \textit{w/o AT} & 7.394 & 14.493 & 2.388 & 4.852 & 3.418 & 7.709 \\
    \bottomrule
    \end{tabular}
    \caption{Ablation Study with modules removed from KDSR, in three datasets. AT denotes the asynchronous training strategy.}
    \label{tab:ablation}
  \end{minipage}%
  \hspace{2mm}
  \begin{minipage}[t]{0.39\textwidth}
    \vspace{0pt}
    \centering
    \scriptsize
    \tabcolsep=0.9mm
    \begin{tabular}{c|cc|cc}
    \toprule
    \multirow{2}{*}{$\xi$} & \multicolumn{2}{|c}{\textbf{VQ}} & \multicolumn{2}{|c}{\textbf{$k$-Means}}  \\
     & H@5 & H@20 & H@5 & H@20  \\
    \midrule
    \textit{Euc} & 7.121 & \textit{14.657} & \textit{7.201} & 14.631 \\
    \textit{Man} & \textit{7.174} & \textit{14.569} & 7.165 & 14.537 \\
    \textit{Dot} & 7.117 & \textit{14.519} & \textit{7.118} & 14.507 \\
    \textit{Cos} & \underline{\textit{7.245}} & \underline{\textit{14.752}} & 7.224 & 14.743 \\
    \bottomrule
    \end{tabular}
    \caption{Quantization Functions over Beauty datasets. \textit{Italics} indicate the better, and \underline{underline} indicates the best.}
    \label{tab:correlation}
  \end{minipage}
  \vspace{5mm}
\end{table}

\vspace{0.5em}
\noindent \textbf{Correlation Distillation Study.}
Capturing the correlation supervision signal is pivotal in our method, and we examine this through two lenses: quantization methods (Vector Quantization and $k$-Means) and scoring functions (Euclidean/ Manhattan distance, and Dot/Cosine similarity).

In evaluations using H@5 and 20 (Table~\ref{tab:correlation}), VQ and $k$-Means have distinct performances.
While both are unsupervised quantization approaches, VQ notably surpasses $k$-Means in noise resilience. We attribute this to VQ's adaptable codebook training from scratch, contrasting with $k$-Means' reliance on random initial centroid placements.
The inherent refined VQ representations thrive in a wider range of contexts, evidenced by VQ's leading H@20 scores. Conversely, in scenarios like H@5 where precision is crucial, $k$-Means may occasionally (\textit{Euc} and \textit{Dot}) yield more precise outcomes.
Meanwhile, cosine similarity emerges as the most compatible scoring function for both VQ and \(k\)-Means. 
These results may arise from cosine similarity's resilience with sparse data, as distance-based metrics can be swayed by noise. 
Furthermore, the distinction between Cosine and Dot arises from the dot product's sensitivity to vector magnitudes, making it prone to distortions from noise and outliers. Conversely, cosine similarity is normalized, which presents more consistent and robust results.

\begin{figure}[t!]
    \centering
    \setlength{\belowcaptionskip}{-2em}
    \subfigure{
		\includegraphics[width=0.235\linewidth]{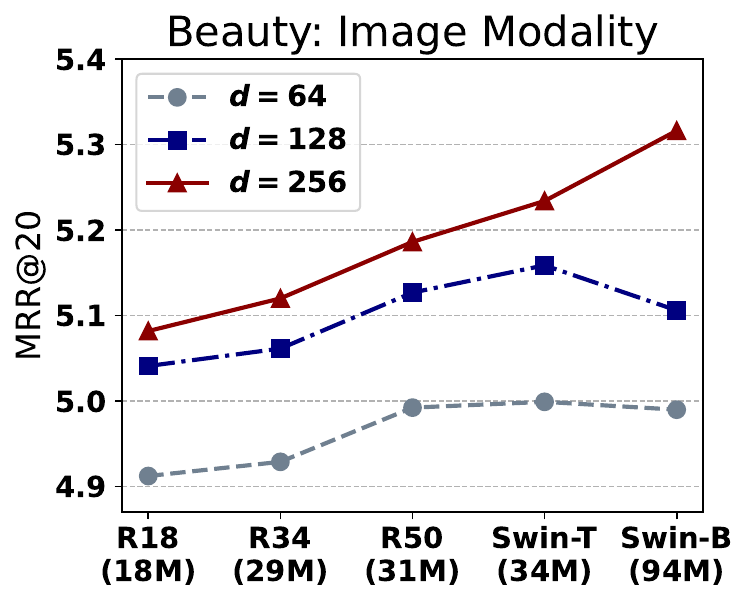}}
    \subfigure{
		\includegraphics[width=0.235\linewidth]{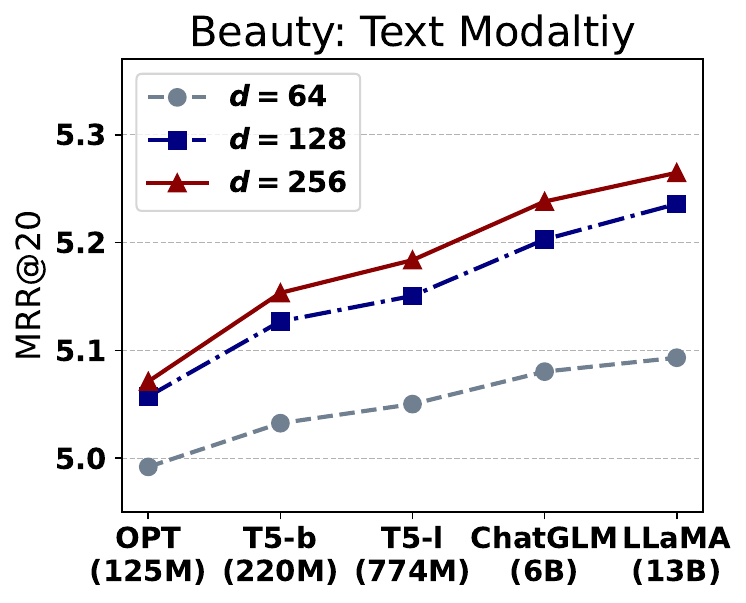}}
    \subfigure{
		\includegraphics[width=0.235\linewidth]{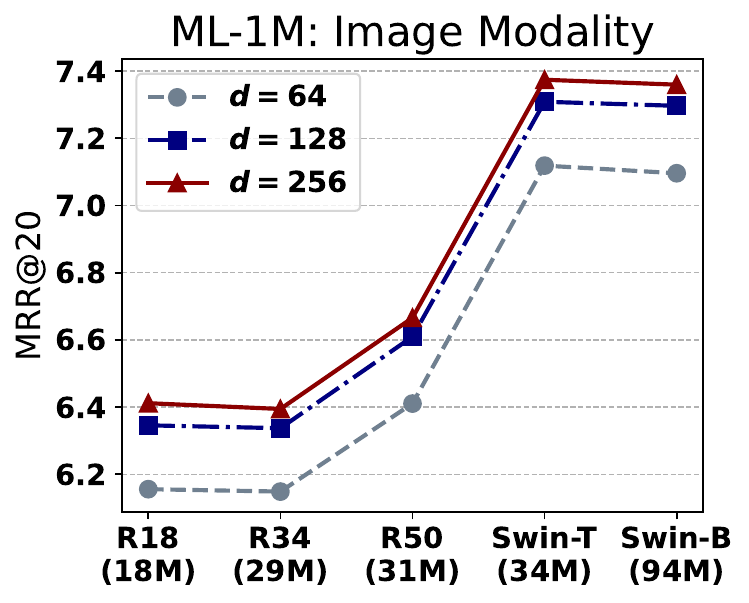}}
    \subfigure{
		\includegraphics[width=0.235\linewidth]{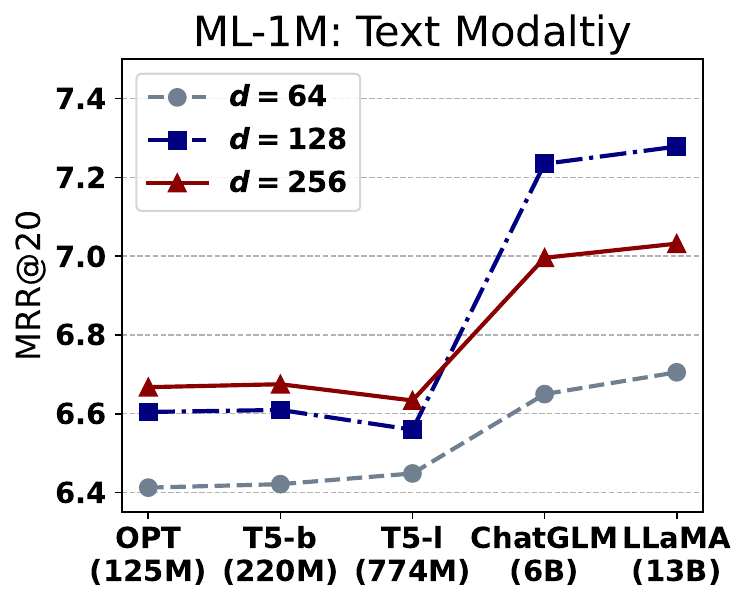}}
    \caption{Evaluation on two datasets equipped with MMSR backbone and different modality encoders (image and text) with various parameter sizes. R* indicates the ResNet \cite{he2016deep} with different sizes. Swin-T/B are the transformer-based models \cite{liu2021swin} in two sizes. T5-b/l indicates the base and large version of T5 \cite{raffel2020exploring}. ChatGLM \cite{zeng2023glm130b} and LLaMA-13B \cite{touvron2023llama} are two recently introduced LLMs.}
    \label{fig:encoder}
\end{figure}

\subsection{Influence of Modality Encoder Sophistication}
\label{sec:exp-3}
As modality encoders become more sophisticated in terms of parameters, they become better equipped to capture nuanced modality features. To assess our method's proficiency in grasping correlations amid these refined features, we executed controlled experiments (in Figure~\ref{fig:encoder}), tweaking the encoder for each modality.

Overall, for both long (ML-1M) and short (Beauty) sequence tasks, modality encoders with richer parameters extract more information, generating more accurate dissected correlations resulting in improved embedding learning.
If a modality's embedding dimension ($d$) is too small, it might struggle to both represent the original modality and support sequential modeling. For instance, in the Beauty dataset discussing image encoders, when the embedding dimension $d= 64$ or 128, overall performance declines slightly as the encoder parameter size increases. But this does not happen when $d=256$.
 The optimal modality embedding size varies across scenarios. In ML-1M, for text representations extracted by LLM (i.e., ChatGLM or LLaMA), the embedding dimension $d=256$ performs worse than $d=128$. This is likely because longer sequences will cause an exponential increase in item relation pairs. 
This will lead to more numerous correlation supervisions which may result in the embedding overfitting to the CDL task, leading to suboptimal outcomes.




\section{Conclusion}




As encoded modality features remain underleveraged in two-stage SRs, and current modality encoders grow in complexity, we introduce a lightweight adaptation strategy.
Concerning modality forgetting during SR training, we leverage self-supervised knowledge distillation.
It first comprehensively distills both fine- and coarse-grained correlation signals from encoded modality representations, and then enhances embedding learning through these signal supervision. 
Experiments underscore that while holistic correlations augment accuracy, dissected correlations enhance diversity.  Together they ensure a balanced recommendation.

In this study, we present an efficient approach to distill knowledge from the output of the modality encoder's last layer, leaving room for integrating deeper encoder layers in future research. On the technical front, our current distilled knowledge is in the format of triplets, i.e., \( (v_i,v_j,r_{ij}) \). Further adopting structured distillation formats, like graphs, may also be a compelling next step. 
Additionally, while we emphasize intra-modality supervision, we believe it is promising to study joint supervision across varied modality types, enhancing multi-modal SR tasks.




\bibliographystyle{splncs04}
\bibliography{mybibliography}

\appendix

\end{document}